\documentclass[fleqn,usenatbib]{mnras}

\usepackage[T1]{fontenc}
\usepackage{ae,aecompl}

\usepackage{graphicx}	
\usepackage{amsmath}	
\usepackage{amssymb}	
\usepackage{natbib}
\usepackage[varg]{txfonts}

\usepackage{mathrsfs} 
    
\usepackage{xspace}
\newcommand{\msun}{\ensuremath{\mathrm{M}_{\odot}}\xspace}

\newcommand{\ttau}{\ensuremath{T(\tau)}\xspace}

\newcommand{\Kp}{\textit{Kepler}}
\newcommand{\rd}{\color{black}}
\newcommand{\figs}{./}

\title[Modelling Linewiths of RGB stars in NGC 6819]{Modelling linewidths of \textit{Kepler} red giants in NGC 6819}

\author[M. J. Aarslev et al.]{
Magnus J. Aarslev\thanks{Contact e-mail: mja@phys.au.dk},
G{\"u}nter Houdek,
Rasmus Handberg,
\newauthor J{\o}rgen Christensen-Dalsgaard
\\
Stellar Astrophysics Centre (SAC), Department of Physics and Astronomy, Aarhus University, Denmark}

\date{Accepted XXX. Received YYY; in original form ZZZ}

\pubyear{2016}

\begin{document}
\label{firstpage}
\pagerange{\pageref{firstpage}--\pageref{lastpage}}
\maketitle

\begin{abstract}
We present a comparison between theoretical, frequency-dependent, damping rates and linewidths of radial-mode oscillations in red-giant stars located in the open cluster NGC 6819. The calculations adopt a time-dependent non-local convection model, with the turbulent pressure profile being calibrated to results of 3D hydrodynamical simulations of stellar atmospheres. The linewidths are obtained from extensive peakbagging of {\Kp} lightcurves. These observational results are of unprecedented quality owing to the long continuous observations by {\Kp}. The uniqueness of the {\Kp} mission also means that, for asteroseismic properties, this is the best data that will be available for a long time to come. We therefore take great care in modelling nine RGB stars in NGC 6819 using information from 3D simulations to obtain realistic temperature stratifications and calibrated turbulent pressure profiles.
Our modelled damping rates reproduce well the {\Kp} observations, including the characteristic depression in the linewidths around the frequency of maximum oscillation power. Furthermore, we thoroughly test the sensitivity of the calculated damping rates to changes in the parameters of the nonlocal convection model.
\end{abstract}




\section{Introduction}
\label{sec:intro}
Solar-like oscillations are characterized by stochastic excitation of modes, when sound waves resonate with convective motion of the gas, e.g. \cite{Houdek1999AAp}. The amplitude of a stochastically driven mode decays with a lifetime $\tau = \eta^{-1}$, with $\eta$ being the damping rate in units of angular frequency $\rm s^{-1}$. The damping rate is frequency dependent and is related to the observed linewidth, $\Gamma$, by the relation $\Gamma = \eta/\pi$. Here the linewidth, measured in units of cyclic frequency $\rm Hz$, is the full width at half maximum (FWHM) of a Lorentzian obtained from fitting a peak in the frequency power spectrum of an observed light curve. For computational purposes stellar oscillations are most often assumed to be adiabatic. Even though this is in some cases sufficient, e.g. fitting observed frequencies, when complemented by frequency corrections, it is inadequate for studying and understanding the complete physical picture including mode damping. Furthermore, the stratification in the surface layers cannot be adequately modelled in a local formulation of convection customarily used in 1D stellar models.

The complexity of non-adiabatic pulsations and their coupling to the convection has posed many problems since the field's inception and still does. The main problem lies in our, so far, limited understanding of the interaction between convection and pulsations. However, several important steps forward have already been taken, and several recent reviews on the topic exist \citep[see for example][]{houdek2015LRSP,samadi2015proceedings}. The case of solar pulsational stability has been studied in detail both theoretically \citep{Balmforth1992part1MNRAS} and observationally \citep{chaplin1997MNRAS,komm2000ApJ}, while the space missions CoRoT \citep{baglin2006} and {\Kp} \citep{boruckiSci2010,borucki2016RPPh} have provided high-quality seismic data for stars of different flavours against which we can test models and further our understanding of stellar pulsations. \cite{appourchaux2014AA} analysed oscillation mode linewidths for a number of {\Kp} main-sequence solar-like stars and found interesting relationships between linewidths, frequencies and effective temperatures. 
Using CoRoT observations \cite{samadi2012AA} showed that non-adiabatic effects are present and non-negligible in red-giant stars. \cite{HoudekGough2002MNRAS} modelled damping rates and velocity amplitudes of the red giant $\xi$ Hydrae, while \cite{dupret2009AA} computed theoretical amplitudes, lifetimes and heights in the frequency power spectrum of oscillation modes at different stages of red giant evolution, including the phase of core helium burning, and \cite{grosjean2014AA} computed synthetic power spectra for mixed modes in red giants. \cite{belkacem2012AA} were able to reproduce observed $\Gamma$ vs $T_{\rm eff}$ across the HR-diagram including both main-sequence as well as red-giant stars. However, calculations of frequency-dependent damping rates for red-giant stars have so far not been able to survive comparisons with observations. \cite{handberg2017MNRAS}, hereafter referred to as H17, obtained precise frequencies and linewidths for a sample of red giants in NGC 6819 by means of extensive, careful peak bagging. Here, we compute frequency-dependent damping rates for a selection of red-giant-branch (RGB) stars in the H17 sample. This is done via a non-adiabatic stability calculation \citep{Houdek1999AAp} for which we obtain the convective fluxes from a non-local, time-dependent convection model \citep{gough1977ApJ,gough1977LNP} partly calibrated through 3D convection simulations \citep{rt2013apj}.

\section{The stellar sample NGC 6819}
NGC 6819 is a solar-metallicity open star cluster located in the {\Kp} field. H17 presented the first extensive peakbagging effort on {\Kp} light curves of evolved red giants (RGB as well as clump stars) in NGC 6819. Preliminary results were presented in \cite{handberg2016AN}. The results include not only precise individual frequencies but also their corresponding linewidths $\Gamma$, as well as masses, effective temperatures and large frequency separations needed to model the stars. Here we present a successful attempt at reproducing observed linewidths $\Gamma$ of radial pressure-mode (p) oscillations by means of theoretical calculations for nine red giants in NGC 6819. This has previously been done successfully for main-sequence solar-like stars, \citep[e.g.][]{houdek2017arXiv}. Applying the same technique to obtain frequency-dependent damping rates for red giants has so far been uncharted territory.

The mean RGB mass in NGC 6819 is $1.61\pm 0.02\rm M_{\odot}$ and the metallicity is practically solar. Recent analysis of high resolution, high signal-to-noise spectroscopic data of KIC 5024327 in NGC 6819 yielded $\left[ \rm Fe / \rm H \right] = -0.02 \pm 0.10$ \citep{slumstrup2017AA}, and in addition gave a surface gravity value of $\log g=2.52$ in agreement with the asteroseismic value of $2.546$ \citep{corsaro2012ApJ}. This is part of what makes NGC 6819 particularly interesting, especially from a modelling viewpoint, because it enables the consistent use of realistic \ttau relations calculated from a solar-metallicity grid of 3D stellar atmosphere models (see Section \ref{sec:methods}). The age of the cluster is 2.25 Gyr according to several studies including those by H17 and \cite{bedin2015MNRAS}. All stars considered here have been labelled as single members of NGC 6819.

The peak bagging of {\Kp} lightcurves by H17 resulted in 5-6 oscillation modes for each of the spherical degrees $l=0,1,2$. We restrict the present calculation of frequency-dependent damping rates to radial modes. The radial orders $n$ with which we compare are typically between $5$ and $15$. For the Sun the energetics of low-degree ($l \lesssim 100$) p-mode oscillations are almost independent of $l$ \citep{jcd1982MNRAS}. Unfortunately this does not apply to our analysis. As stated above we do have independent observed linewidths for oscillations modes other than $l=0$ modes. But most likely these are not pure p modes but have a mixed character, behaving as gravity (g) modes in the deep interior. Our results are therefore only valid for $l=0$ modes.

\section{Numerical Procedures}
\label{sec:methods}
\subsection{Evolution models with 3D atmosphere $T(\tau)$ relations}
We calculate frequency-dependent damping rates using a non-adiabatic pulsation code, which includes perturbations to the convective heat flux and turbulent pressure. These are obtained from a 1D envelope model based on a nonlocal, time-dependent convection model incorporating turbulent pressure, initially developed by \cite{gough1977LNP, gough1977ApJ}. The same convection formulation is used consistently in the pulsation and equilibrium codes and contains parameters controlling the degree of nonlocality, discussed in Section \ref{sec:em}, as well as a nonlocal mixing-length parameter $\alpha_{\rm NL}$. The mixing length is typically calibrated so that the model yields a desired depth of the convective envelope (matching either a model or a value extracted from observations). However, for the red giants considered here, the radius at the bottom of the convective zone $r_{\rm BCZ}$, as given by evolutionary models, is very small, ranging approximately from $0.05R$ to $0.1R$, where $R$ is the radius of the star. In principle $r_{\rm BCZ}$ could be determined from observations via acoustic glitch signatures from the base of the convective envelope. This can be difficult but has been done for main-sequence stars, e.g., by \cite{Mazumdar2014ApJ, verma2017ApJ}. For red giants the amplitudes of these glitch signatures are tiny, which complicates such a determination of $r_{\rm BCZ}$. Because $r_{\rm BCZ}$ is then not reliably known we instead adjust $\alpha_{\rm NL}$ so that the the sound speed profile, calculated as $c_{\rm s}^2 = \gamma_1 (p_{\rm g}/\rho)$, agrees with that of a full structure model in the deep interiors, where $p_{\rm g}$ is gas pressure, $\rho$ is density, and $\gamma_1 = \left( \partial\ln p_{\rm g} / \partial\ln\rho \right)_{\rm ad}$ with "$\rm{ad}$" denoting an adiabatic derivative. The reason we do not readily adjust $\alpha_{\rm NL}$ to adopt $r_{\rm BCZ}$ from a full structure model is the presence of hydrogen-burning shells very close to the bottom of the convective envelope. This affects slightly the structure in those deep layers in a manner that is not reproduced if the same $r_{\rm BCZ}$ location is retained in the envelope model, which by construction does not contain energy production. Being interested in the seismic properties of the model, we allow for small deviations from $r_{\rm BCZ}$ in order to have matching sound speed profiles. The values we obtain for $\alpha_{\rm NL}$ are given in Table \ref{tab:models}\footnote{We want to stress that the mixing-length parameter $\alpha_{\rm NL}$ in the nonlocal convection model is not comparable to that of local theories such as the B{\"o}hm-Vitense formulation. Therefore the values of $\alpha_{\rm NL}$ given here should not be used in any other context.}.

The appropriate stellar structures are obtained by evolving models with the Aarhus STellar Evolution Code (ASTEC) \citep{astec2008}. Adiabatic oscillation frequencies are calculated with the Aarhus adiabatic oscillation package (ADIPLS) \citep{adipls2008} in order to match the mass, large frequency separation $\Delta\nu$ (the mean frequency spacing between two consecutive radial modes), and effective temperature $T_{\rm eff}$. The only free parameter that we adjust is the local mixing-length parameter $\alpha_{\rm MLT}$ \citep{bohm-vitense1958}. The values of $\alpha_{\rm MLT}$ are calibrated such as to match $T_{\rm eff}$ and $\Delta\nu$ determined by H17, while the masses are fixed by seismic scaling relations.

At the surface boundary of the models we construct the temperature structure with realistic $T(\tau)$ relations, between temperature $T$ and optical depth $\tau$, calculated from a grid of 3D simulations of radiation-coupled hydrodynamics in stellar atmospheres by \cite{rt2013apj}. The calculation and implementation details are described in \cite{rt2014ttau}. In the stellar evolution models described above, the metallicity is chosen to be consistent with the simulations ($X=0.737$ and $Z=0.018$). The simulation grid consists of 37 atmospheres irregularly spaced in $\log g$ and $T_{\rm eff}$ and covers the full main-sequence and RGB evolution for $\log g \gtrsim 2.5$ of stars approximately in the mass range $0.7$-$1.4$\msun. The appropriate $T(\tau)$ relation can then be interpolated and extracted at every time step in the evolutionary calculation. However, since the RGB stars in NGC 6819 are more massive than $1.4$ \msun, their $T_{\rm eff}$ can exceed that of the \ttau tables provided by the grid of simulations. Consequently, $T(\tau)$ relations cannot be interpolated reliably during that phase. For this reason we use just the $T(\tau)$ relation calculated from the solar simulation. Aarslev et al. (in prep.) show that there is almost no difference between this approach and continuously updating the $T(\tau)$ relation during evolution. Both approaches differ from, and should be an improvement to, traditional semi-empirical $T(\tau)$ relations. For the 1D nonlocal envelope model and stability computations the stars are well within the tables of \ttau relations, in which we interpolate to get the best possible boundary conditions.

\begin{figure}
\includegraphics[width=\columnwidth]{\figs/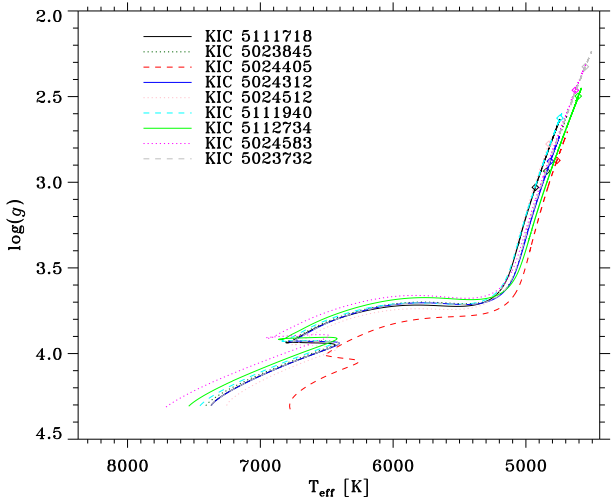}
	\caption{Evolutionary tracks for the stars that have been modelled. The diamond-shaped points show where in the Kiel diagram the observed stars are located.}
	\label{fig:HR}
\end{figure}

\begin{figure}
\includegraphics[width=\columnwidth]{\figs/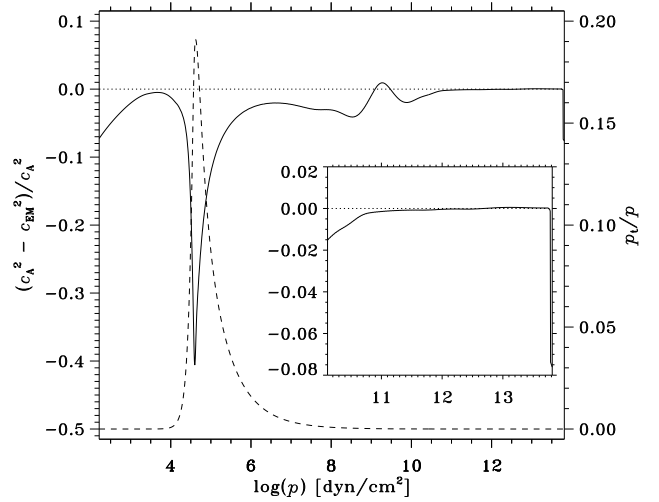}
\caption{The solid curve shows the difference in sound speed profile between the ASTEC evolutionary model ($c_{\rm A}$) and the nonlocal envelope model ($c_{\rm EM}$) of KIC 5111718. The dashed curve shows the ratio of turbulent pressure to total pressure $p_{\rm t}/p$.}
	\label{fig:calistruct5111718}
\end{figure}

Figure \ref{fig:HR} shows the evolutionary tracks for the modelled red giants. The diamond-shaped points show where the observed stars are located. Figure \ref{fig:calistruct5111718} illustrates an example of the $\alpha_{\rm NL}$ calibration. The dashed line shows the ratio of turbulent pressure $p_{\rm t}$ to total pressure $p$, the peak of which occurs near the maximum of the super-adiabatic temperature gradient. This is where the ASTEC evolutionary models differ the most from the nonlocal envelope models, as they should. Deep below the surface the models are almost identical. The horizontal axis on Figure \ref{fig:calistruct5111718} shows the total pressure $p$ of the nonlocal envelope model. The difference between the models in the deep interior is due to hydrogen shell burning in the ASTEC evolutionary model.

\subsection{Details of nonlocal convection model}
\label{sec:em}
Here we give a short summary of the main equations of the convection formulation used in the nonlocal envelope and pulsation models but refer the reader to \cite{gough1977LNP}, \cite{gough1977ApJ}, \cite{Balmforth1992part1MNRAS} and \cite{houdek1996phd} for the full explanation as well as to the review by \cite{houdek2015LRSP}. The theory is based on a generalized mixing-length formulation proposed by \cite{spiegel1963ApJ}, where one considers for the turbulent convective elements a distribution function in six-dimensional phase space, which gives rise to a transfer equation describing the conservation of eddies. \cite{gough1977ApJ} applied this formulation to his time-dependent mixing-length model to obtain a nonlocal generalization which we adopt here. In this generalization the nonlocal convective heat flux $F_{\rm c}$ is obtained by considering its local solutions $\mathscr{F}_{\rm c}$ as source functions weighted by a smoothing kernel $\mathscr{K}$, leading to
\begin{equation}
\label{eq:integratelocal}
F_{\rm c} = \int_{-\infty}^{\infty}{\mathscr{F}_{\rm c}(\xi_0) \mathscr{K}(\xi,\xi_0) \rm{d}\xi_0  } ,
\end{equation}
where the exact kernel ${\mathscr{K}}$ is the second exponential integral. \cite{gough1977ApJ} approximates this kernel as
\begin{equation}
\label{eq:kernel}
\mathscr{K}(\xi,\xi_0,a) \simeq \frac{1}{2}a{\rm e}^{-a|\xi-\xi_0|} ,
\end{equation}
with $a$ being a dimensionless parameter controlling the degree of nonlocality of the convective flux, and $\xi$ is the dimensionless vertical displacement, ${\rm d}\xi = -{\rm d} z/\ell$, of an element from its initial position $\xi_0$ ($\ell$ is the local mixing length). With the approximated kernel $\mathscr{K}$ in Eq.~\eqref{eq:kernel}, the nonlocal heat flux in Eq.~\eqref{eq:integratelocal} is the solution of the second-order differential equation
\begin{equation}
\label{eq:diffeq_flux}
\frac{1}{a^2}\frac{{\rm d}^2 F_{\rm c}}{{\rm d}\xi^2} = F_{\rm c} - \mathscr{F}_{\rm c}  .
\end{equation}
Low values of $a$ provide strongly nonlocal solutions, while large values result in localized solutions. 

A similar expression is obtained for the momentum flux, or turbulent pressure $p_{\rm t}$
\begin{equation}
\label{eq:diffeq_pt}
\frac{1}{c^2}\frac{{\rm d}^2 p_{\rm t}}{{\rm d}\xi^2} = p_{\rm t} - \mathscr{P}_{\rm t}  ,
\end{equation}
where $\mathscr{P}_{\rm t}$ is the local turbulent pressure, and the dimensionless parameter $c$ has a similar meaning as $a$ in eq. \eqref{eq:diffeq_flux}\footnote{Note that \citep{Balmforth1992part1MNRAS} uses a single parameter $a$ for the nonlocality of both the convective flux and turbulent pressure.}.
Furthermore, in order to account for the comparatively short trajectories of the convective elements described by the local source function, \cite{spiegel1963ApJ} suggested to replace the super-adiabatic gradient $\beta$ in the local formulation,
\begin{equation}
\label{eq:superad-gradient}
\beta := -\frac{{\rm d}T}{{\rm d}r} + \frac{\delta}{\rho c_p}\frac{{\rm d}p_{\rm g}}{{\rm d}r} = -\frac{T}{H_ p}\left( \nabla - \nabla_{\rm ad} \right),
\end{equation}
by the nonlocal quantity $\mathscr{B}$ obtained from the solution of
\begin{equation}
\label{eq:diffeq_grad}
\frac{1}{b^2}\frac{{\rm d}^2 \mathscr{B}}{{\rm d}\xi^2} = \mathscr{B} - \beta
\end{equation}
with $b$ being analogous to $a$ and $c$ introduced above. In expression \eqref{eq:superad-gradient} $r$ is radius, $p_{\rm g}$ is the gas pressure, $\delta = -(\partial \ln\rho / \partial\ln T)_{p_{\rm g}}$ is the isobaric expansion coefficient, $\rho$ is the density, $c_p$ the specific heat at constant gas pressure, and $H_p$ is the pressure scale height. The last quantity is the superadiabatic gradient $\nabla - \nabla_{\rm ad}$, where $\nabla := ({\rm d}\ln T/{\rm d}\ln p)$ is the temperature gradient and $\nabla_{\rm ad}$ is the adiabatic temperature gradient.

The convection parameters $a$, $b$, and $c$ are in principle free parameters, although theoretical values were suggested \citep[e.g.,][and references therein]{gough1977LNP}. However, $c$ can be fairly tightly constrained; the main impact of $c$ on the model is the maximum value of $p_{\rm t}/p$ in the superadiabatic boundary layer, which can be inferred from 3D convection simulations (Figure 6 in \cite{rt2013apj} shows this maximum value across the HR-diagram). We use this to calibrate $c$, which is typically in the order of $c^2 \approx 150$, whereas $a$ and $b$, adjusted to achieve the best agreement between damping rates and observed linewidths, attain a much broader range of values, $b$ varying the most between different stars. It speaks to the strength of the model that the stellar depth at which the $p_{\rm t}/p$ peaks matches that of 3D convection simulations \citep{houdek2017MNRAS}. Because $a$ and $b$ are not easily determined, it is of special interest to study how they affect both the physical structure of the model as well as the resulting damping rates. This is discussed in detail in Section \ref{sec:emtest} along with the effect of the shape parameter $\Phi$, describing the anisotropy of the convective velocity field, defined as
\begin{equation}
\label{eq:Phi}
\Phi := \frac{\langle u^2 + v^2 + w^2\rangle}{\langle w^2\rangle}  ,
\end{equation}
where angular brackets denote horizontal mean values, $u$ and $v$ are horizontal velocities, while $w$ is the vertical component of the turbulent velocity field so that isotropic turbulence corresponds to $\Phi = 3$. The parameter enters into the equations of motion in Gough's convection model as a virtual increase of inertia of the fluid elements. It will be described in more detail in Section \ref{sec:emtest_results}.

In Section \ref{sec:intro} we introduced the relation between damping rates and linewidths, $\Gamma = \eta/\pi$. It should be noted that this relation assumes that $\eta$ includes all possible contributions to mode damping. However, we limit ourselves to the three primary contributions, namely the modulation by the pulsations of the convective heat flux $F_{\rm c}$ and momentum flux $p_{\rm t}$, respectively, as well as radiative damping. Consequently, we omit contributions to $\eta$ from incoherent scattering at the super-adiabatic boundary layer as well as energy lost by waves being transmitted into the atmosphere (e.g. \cite{Houdek1999AAp,BalmforthGough1990ApJ} and references therein).

In both the nonlocal envelope and pulsation code, we treat radiative transfer in the generalized Eddington approximation \citep{UnnoSpiegel1966PASJ}. We use the OPAL opacities \citep{IglesiasRogers1996ApJ} with additional low-temperature opacities obtained from \cite{kurucz1991ASIC}. We use an equation of state (EOS) including a detailed treatment of the ionization of C, N, and O, as well as the first ionization of the next seven most abundant elements \citep{eggleton1973}. In order to be consistent with the 3D convection simulations, we adopt in all three aforementioned codes the same hydrogen and heavy element abundances, which are $X=0.737$ and $Z=0.018$ respectively \citep{rt2014alpha}.

\section{Tests of convection-model parameters}
\label{sec:emtest}
We calibrate the nonlocal convection parameter $c$ via 3D convection simulations, while the nonlocal mixing-length parameter $\alpha_{\rm NL}$ is calibrated to obtain the deep structure of the corresponding evolutionary models. We use the observed linewidths to calibrate $a$, $b$, and $\Phi$, although the values we use for $\Phi$ are similar to what is found in 3D simulations just below the superadiabatic peak, where $\Phi$ is almost constant with increasing stellar depth. We emphasize that the values of the parameters $a$, $b$, $c$, and $\Phi$ are the same in the nonlocal envelope as well as in the pulsation computations. As such it is interesting to see how these parameters separately affect both the structure and the damping rates of the models. As we show in Section \ref{sec:results}, our model computations are in most cases able to closely reproduce observed linewidths.

\cite{Balmforth1992part1MNRAS} studied the effect of varying the nonlocal convection parameters on the stability solutions of solar models. Here, we basically follow Balmforth's approach and investigate the sensitivity of the model solutions to changes in convection parameters by vaying $a$ and $b$ independently while keeping $\alpha_{\rm NL}$ and $\Phi$ constant. Besides concerning RGB stars, our test is in some aspects a little different: First of all our model includes an additional parameter $c$, which is subject to sensitivity tests as well, but is calibrated to 3D simulation results. Furthermore, we also test for effects of varying $\Phi$. Regarding the mixing-length parameter $\alpha_{\rm NL}$, if it is kept constant the size of the convective envelope changes upon varying the other convection parameters. Therefore we opt for recalibrating $\alpha_{\rm NL}$ to retain the depth of the convective envelope, which is then constant throughout all the tests. One might argue that it will be difficult to disentangle the effects of changing either $\alpha_{\rm NL}$ or, say, $a$. However, we did do the same tests with constant $\alpha_{\rm NL}$, thereby letting $r_{\rm BCZ}$ vary. The effect on the resulting linewidths compared to recalibrating $\alpha_{\rm NL}$ to keep $r_{\rm BCZ}$ constant is extremely small, often indistinguishable.

Lastly, we test the impact of convection-parameter changes on both the structure and dynamics of the envelope model as well as the resulting, frequency-dependent linewidths. It should be noted that the values of the convection model parameters are not directly comparable to \cite{Balmforth1992part1MNRAS}, due to improvements in the code both in the numerics as well as the use of updated opacities, EOS and atmospheric models ($\ttau$) from 3D simulations.

\begin{figure}
\includegraphics[width=\columnwidth]{\figs/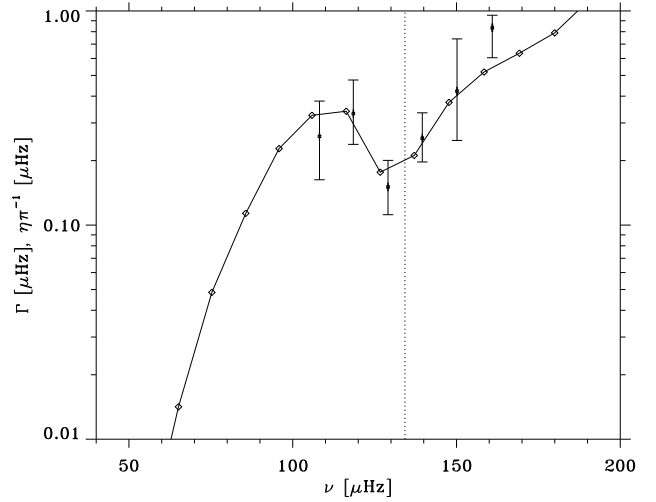}
	\caption{Linewidths of KIC 5111718. Diamonds connected by a full line show model computations as $\Gamma = \eta\pi^{-1}$ while the points with error bars are observed linewidths. The dotted vertical line is drawn at the observed $\nu_{\rm max}$.}
	\label{fig:lw5111718}
\end{figure}

As a reference for the sensitivity tests we use our model of KIC 5111718 presented in Fig.~\ref{fig:lw5111718} with convection parameters $a^2 = 900$, $b^2 = 800$, $c^2 = 120$ and $\Phi = 1.8$. This star is chosen due to the robustness of the models across the relevant parameter spectrum. The particular choice of convection parameters yields the best overall match between computed and observed linewidths. When one convection parameter is varied, the remaining parameters are kept constant at the values given above. The radius at the bottom of the convection zone in the reference model is kept constant during all tests at $r_{\rm BCZ} = 0.105R$.

A note about the velocity anisotropy: 3D stellar atmosphere models show a dependence of $\Phi$ on the depth $z$. Work is ongoing to interpolate the $\Phi\left(z\right)$ structure between 3D atmosphere models (Andreas J{\o}rgensen, personal communication), but so far the sparse population of 3D models in the grid at the RGB stage makes it difficult to do so accurately, motivating our use of a constant $\Phi$. It is, however, already feasible for main-sequence solar-like stars; \cite{houdek2017arXiv} presented the first results of using those 3D simulations to guide the functional form of $\Phi(z)$ in the 1D stability
computations for stars in the LEGACY sample \citep{lund2017ApJ, silvaaguirre2017ApJ}.
\subsection{Results of model tests}
\label{sec:emtest_results}
In this section we discuss the effects of the convection-model parameters on both the linewidths and the structure of the models for the test case of KIC 5111718. Note that for too small values for $a^2$ and $b^2$ (below $450$) we find, similarly to previous studies \citep{Houdek1999AAp,HoudekGough2002MNRAS}, unstable modes, i.e. $\eta < 0$, wherefore these models are excluded in the following. This is no detriment to the analysis as $a$ and $b$ typically need to take on much larger values (see Table~\ref{tab:models}). In all figures the black dashed line shows the reference model, which was found to best reproduce the observed linewidths. The parameter values for this model are given in Table \ref{tab:models}.

\begin{figure}
\includegraphics[width=\columnwidth]{\figs/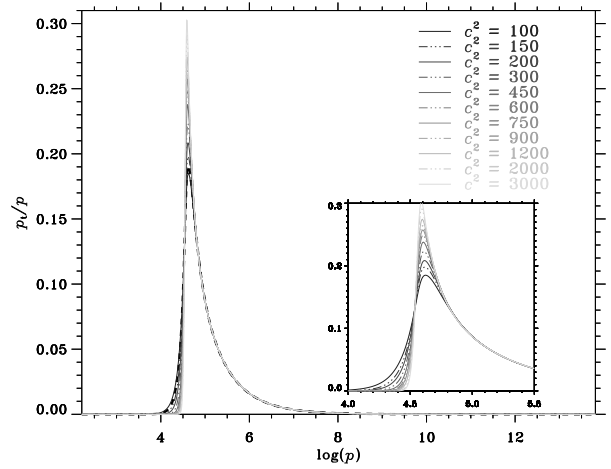}
	\caption{Ratio of turbulent pressure to total pressure from models of KIC 5111718 with different values of the nonlocal convection parameter $c$.}
	\label{fig:ac_pt}

\end{figure}

\begin{figure}
\includegraphics[width=\columnwidth]{\figs/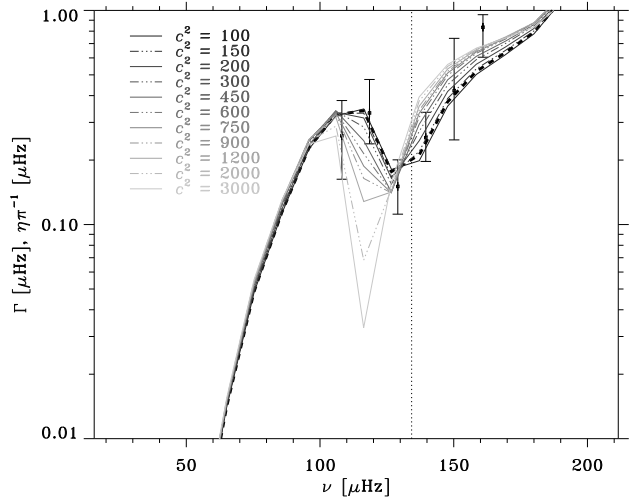}
	\caption{Linewidths of KIC 5111718 models with different values of $c$. The dashed line shows the linewidths of the reference model. Points with error bars are observed linewidths. The dotted vertical line is drawn at the observed $\nu_{\rm max}$.}
	\label{fig:ac_dampingrates5111718}
\end{figure}

\begin{figure}
\includegraphics[width=\columnwidth]{\figs/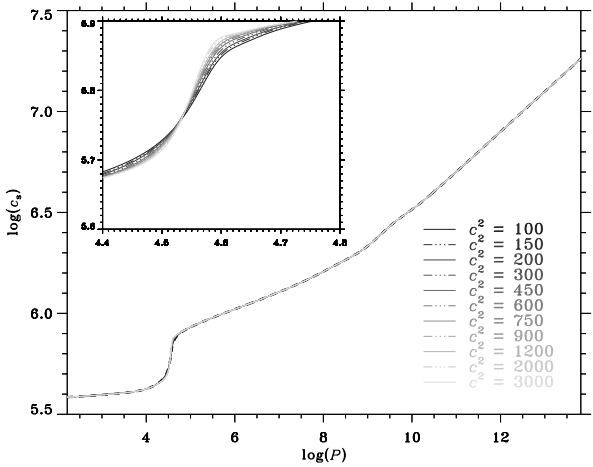}
	\caption{Sound speed from models of KIC 5111718 with different values of the nonlocal convection parameter $c$.}
	\label{fig:ac_cs}
\end{figure}

As described in Section \ref{sec:methods}, we can constrain $c$ much better than the other parameters via the use of 3D hydrodynamics simulations. In order to obtain for $p_{\rm t}/p$ the peak values found in simulations, $c^2$ values range between $110$ and $300$ across our nonlocal models; the specific values for each model are given in Table \ref{tab:models}. Fig.~\ref{fig:ac_pt} shows $p_{\rm t}/p$ for a range of $c$-values. Convection simulation results by \cite{rt2013apj} suggest peak values of $p_{\rm t}/p$ approximately between $0.19$ and $0.22$ for the stars in our sample, thus favouring low $c$ values as per Fig.~\ref{fig:ac_pt}. The impact of $c$ on the linewidth profile for KIC 5111718 is shown in Fig.~\ref{fig:ac_dampingrates5111718}. With increasing $c$ the trough in the linewidths deepens and shifts towards lower frequencies away from the frequency of maximum power $\nu_{\rm max}$. However, a depression of the linewidths around $\nu_{\rm max}$ seems to be a common feature for RGB stars and is also found, but typically not as strongly pronounced, for main-sequence stars \citep{appourchaux2014AA}. The fact that our chosen values of $c$ do not only reproduce the $p_{\rm t}/p$ peak derived from 3D simulations but also the trough in the damping rates at $\nu_{\rm max}$ indicates that models with low $c$ adequately represent the physical conditions of the stars with regards to describing the turbulent-pressure profile in the surface layers. The influence of $c$ on the temperature gradient and convective flux is quite small. But, not surprisingly, the sound-speed profile $c_{\rm s}$ near the surface is sensitive to $c$ as seen in Fig.~\ref{fig:ac_cs}. The sound speed is also affected by $a$ but to a lesser extent and even less for $b$.

\begin{figure}
\includegraphics[width=\columnwidth]{\figs/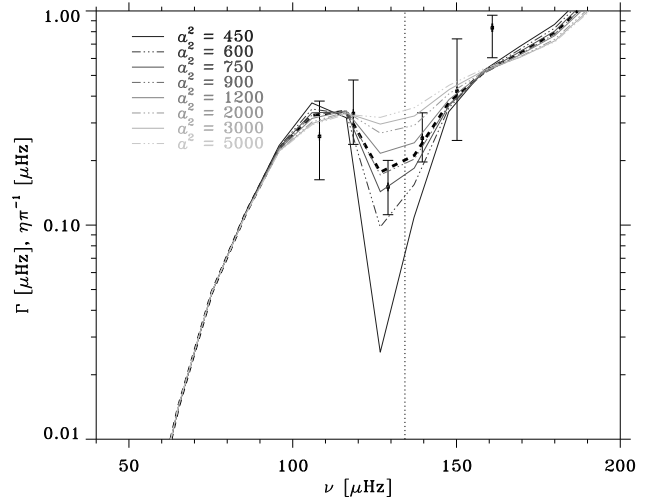}
	\caption{As Fig. \ref{fig:ac_dampingrates5111718} for different values of $a$.}
	\label{fig:aa_dampingrates5111718}
\end{figure}

In Fig.~\ref{fig:aa_dampingrates5111718} we see that $a$ has a rather large impact on the linewidths. The main effect is on the depth of the trough at $\nu_{\rm max}$, the depth of which inceases for decreasing $a$-values while being almost evened out for large $a$. The values listed in Table~\ref{tab:models} suggest that $a$ must, in order for the damping rates to match the observations, increase with evolution to the point where the solutions are practically local. But any conclusion in this regard is prohibited by our limited number of models. We can, however, qualitatively understand this behaviour of the parameter $a$: As the star evolves up the sub-giant branch the convective envelope expands significantly to the point where it constitutes most of the star in terms of volume. In most of the convection zone the pressure scale height, and consequently the mixing length in our description, is so large that it does not make sense to account for contributions from adjacent convective cells when calculating the convective heat flux at a given point. $F_{\rm c}$ is therefore adequately described by a near-local solution.

The fact that the model computations show a clear reduction of the damping-rate depression near $\nu_{\rm max}$ with increasing $a$ indicates that the dynamical effect of varying $a$ is predominantly controlling the degree of the coupling of the (nonlocal) heat flux $F_{\rm c}$ with the pulsationally induced structure variations. The net effect is how the phases between the perturbed heat and radiative fluxes, relative to the density perturbations, are controlled. Increasing $a$, i.e. making $F_{\rm c}$ more local, seems to increase the stabilizing effect of the convective heat-flux perturbation relative to the destabilizing effect of the radiative-flux perturbation on the overall mode stability. 

\begin{figure}
\includegraphics[width=\columnwidth]{\figs/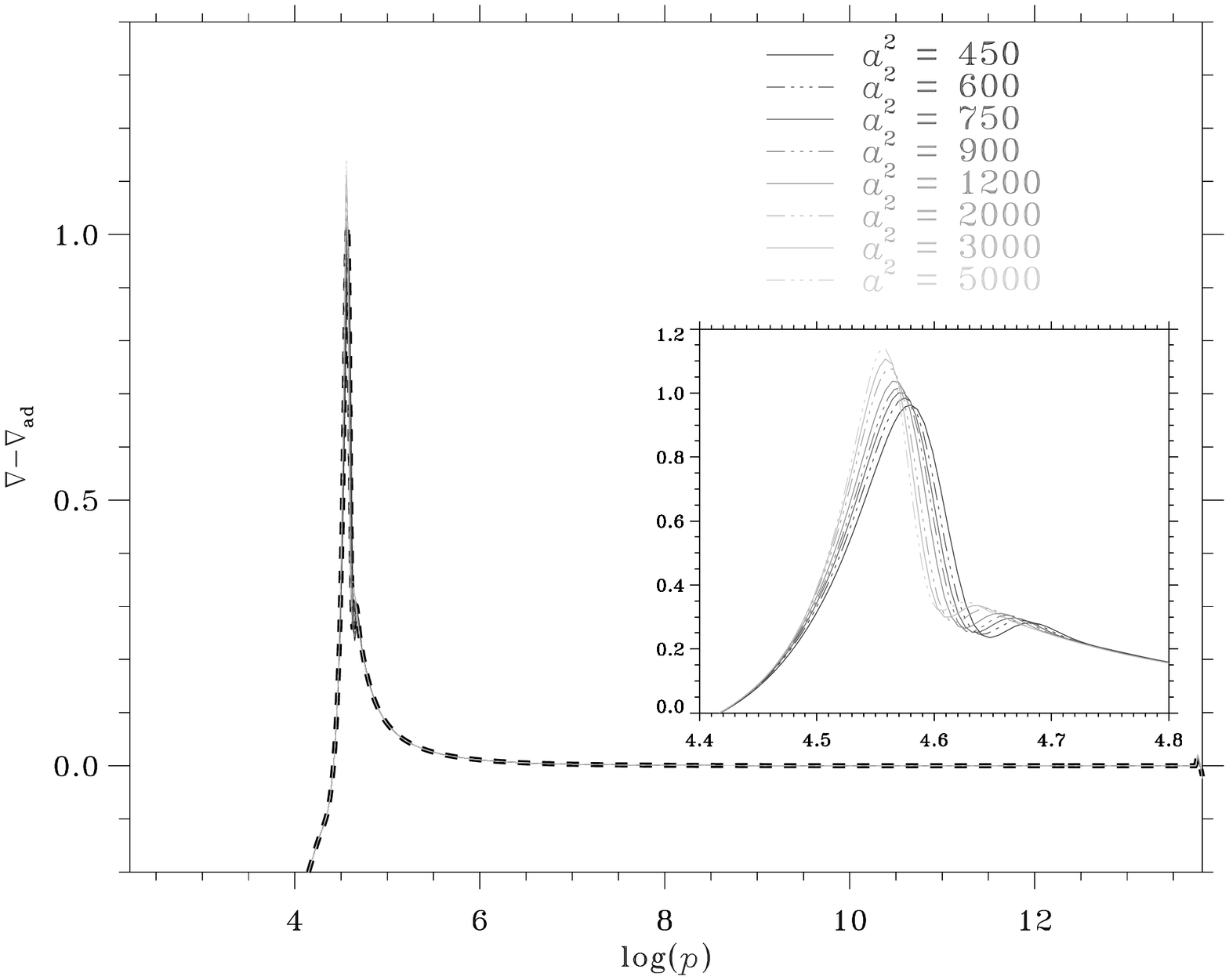}
	\caption{Super-adiabatic gradient from models of KIC 5111718 with different values of the nonlocal convection parameter $a$.}
	\label{fig:aa_grad}
\end{figure}

\begin{figure}
\includegraphics[width=\columnwidth]{\figs/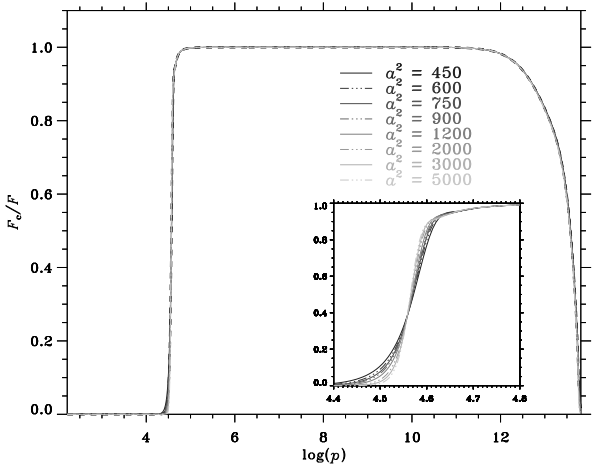}
	\caption{Ratio of convective flux to total flux from models of KIC 5111718 with different values of the nonlocal convection parameter $a$}
	\label{fig:aa_fcon}
\end{figure}

The structural effects of $a$ is shown in Figs~\ref{fig:aa_grad} and \ref{fig:aa_fcon}. The effects on the super-adiabatic temperature gradient, $\nabla-\nabla_{\rm ad}$, and on $F_{\rm c}$ are similar to Balmforth's analysis for the Sun. The turbulent pressure profile is practically unaffected by $a$ (and therefore not shown), as it is, in our models, controlled by $c$ and therefore differs from \cite{Balmforth1992part1MNRAS}. According to 3D simulations, the convective heat flux $F_{\rm c}$ should be negative, albeit very small, in the overshoot region. Contrary to this Fig.~\ref{fig:aa_fcon} shows $F_{\rm c}$ to be still positive and falling off to zero above the super-adiabatic peak (approximately for $\log p < 4.55$). Since we construct the nonlocal flux by integrating local values, it will by design always be positive and confined to the convection zone as determined by the Schwarzschild criterion. It is unlikely that this has a significant effect on the oscillation properties. However, the convection model adopted here includes only the enthalpy flux contribution to $F_{\rm c}$. In reality the convective flux consists of both enthalpy and kinetic energy flux. It is unknown how the latter affects the damping of the acoustic oscillations, but work is ongoing to incorporate it in the convection model.

For $\nabla-\nabla_{\rm ad}$ and $F_{\rm c}/F$ the shifts between models with different values for $a$, seen in Figs~\ref{fig:aa_grad} and \ref{fig:aa_fcon}, are also present at the base of the convection zone. However, at this depth convection does not contribute to the overall stability of acoustic pulsations; we therefore ignore it in the present discussion. We find only slight differences in $\log(T)$ in the superadiabatic layers between models with different $a$, whereas the sensitivity tests of $b$ and $c$ show no effect on the temperature stratification.

\begin{figure}
\includegraphics[width=\columnwidth]{\figs/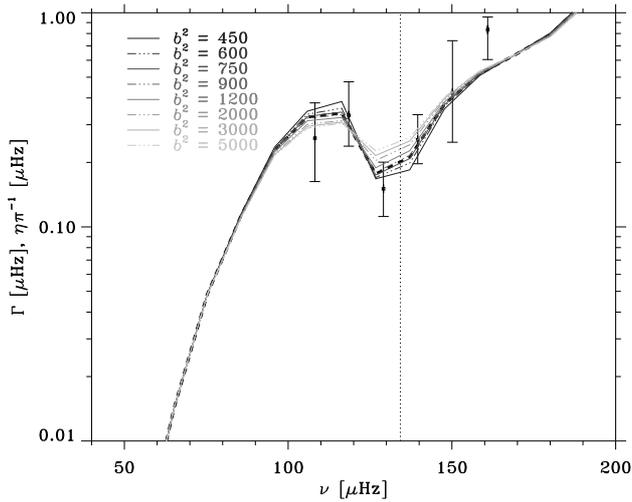}
	\caption{As Fig. \ref{fig:ac_dampingrates5111718} for different values of $b$.}
	\label{fig:ab_dampingrates5111718}
\end{figure}

\begin{figure}
\includegraphics[width=\columnwidth]{\figs/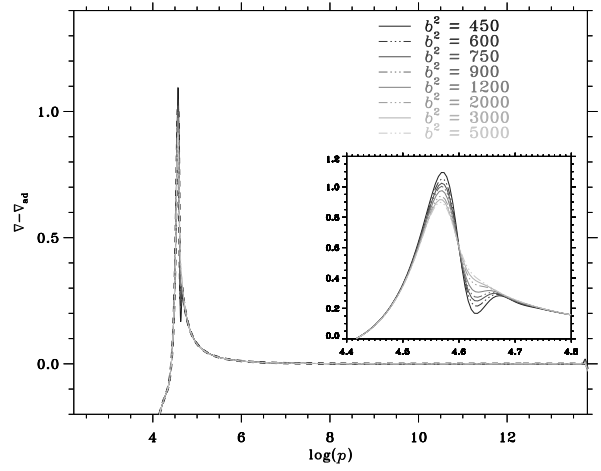}
	\caption{Super-adiabatic gradient from models of KIC 5111718 with different values of the nonlocal convection parameter $b$.}
	\label{fig:ab_grad}
\end{figure}

If we now consider varying $b$, the main effect on the damping rates $\eta$ of increasing the value is to smooth out $\eta$ around $\nu_{\rm max}$ as seen in Fig.~\ref{fig:ab_dampingrates5111718}. Overall the effects on $\eta(\nu)$ seem less drastic compared to varying $a$ or $c$, as depicted in Figs~\ref{fig:ac_dampingrates5111718} or \ref{fig:aa_dampingrates5111718}. Fig.~\ref{fig:ab_grad} shows the effect on $\nabla - \nabla_{\rm ad}$ of varying $b$, which controls the degree to which turbulent fluxes are coupled to the local stratification. The more local the solution the more tightly coupled they are, which suppresses drastic changes in the structure, therefore resulting in lower, and broader, peaks in the super-adiabatic gradient. Smaller $b$-values allow for sharp changes in the structure, which can be brought about by the ionization of hydrogen and result in a small bump in the super-adiabatic gradient just at the bottom of the super-adiabatic layer as seen in Fig.~\ref{fig:ab_grad}. As mentioned by \cite{Balmforth1992part1MNRAS}, extremely low values of $b$ should decrease the temperature gradient enough to bring about a temperature inversion just below the super-adiabatic layer. This feature is seen in laboratory convection; but it is unknown whether or not it can occur in stars. So far, our most realistic representation of convection in stellar atmospheres is provided by 3D simulation, in which no temperature inversions are found. Increasingly local solutions (larger $b$) smooth out this bumpy behaviour as evident in Fig.~\ref{fig:ab_grad}. Compared to \cite{Balmforth1992part1MNRAS}, $b$ has a less significant effect on the convective flux than found for solar models.
The height of the $p_{\rm t}/p$ peak (not shown) is affected only to a minor extent marginally leading to a difference of a little less than $1\%$ in $p_t{\rm }/p$ between the two models furthest apart (highest and lowest $b$).

{\rd Both $a$ and $b$ predominantly affect the linewidths in the depression 
near $\nu_{\rm max}$, with $a$ having the stronger effect.
Thus there is substantial degeneracy between these two parameters,
and a corresponding correlation between the results of fitting them to
observed linewidths, as discussed in Section~\ref{sec:results}.}

\begin{figure}
\includegraphics[width=\columnwidth]{\figs/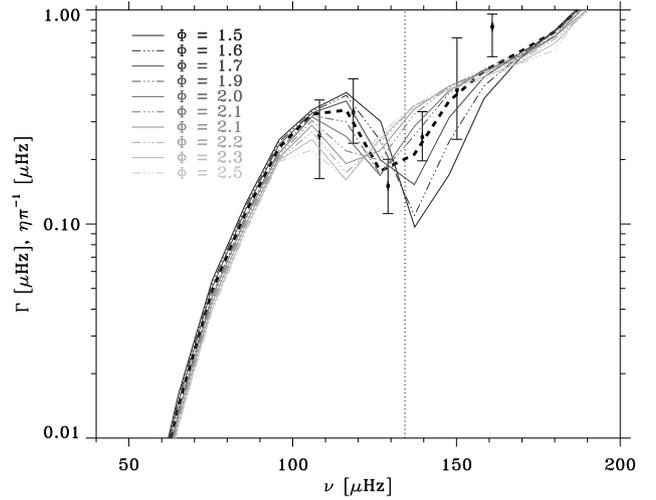}
	\caption{As Fig.~\ref{fig:ac_dampingrates5111718} for different values of $\Phi$. For the reference model (dashed black line) we use $\Phi = 1.8$.}
	\label{fig:ap_dampingrates5111718}
\end{figure}

Fig.~\ref{fig:ap_dampingrates5111718} shows the effect on the linewidths of changing the eddy-shape parameter $\Phi$ (eq. \ref{eq:Phi}). The parameter $\Phi$, which describes the anisotropy of the turbulent velocity field, enters as a multiplicative factor of the inertia term in the fluctuating momentum equation, thereby effectively increasing the inertia of the vertically moving convective eddies as a result of the coupling between vertical and horizontal motion. For a solenoidal turbulent velocity field, it can be related to the shape of the convective eddies in the sense that $\Phi\rightarrow 1$ represents thin, needle-like eddies. Larger $\Phi$ values increase the eddie's inertia, thereby describing the diversion of the vertical motion into horizontal flows as a reduction of the convective efficacy. Moreover, larger $\Phi$ values increase the characteristic timescale of the convection and consequently reduce the frequency at which energy is exchanged most effectively between convection and pulsation. This is demonstrated in Fig.~\ref{fig:ap_dampingrates5111718} by the decrease {\rd with increasing $\Phi$ of the frequency} at which the minimum in the depression of the damping rates is observed. The main effect of varying a depth-independent (constant) $\Phi$ is on the mode dynamics with rather minor effects on the mean structure, mostly due to the convective velocities affecting the turbulent pressure, which can then be mitigated by adjusting $c$.
{\rd Indeed, we note from Figs~\ref{fig:ac_dampingrates5111718}
and \ref{fig:ap_dampingrates5111718} that the effects on the linewidths
of changing $c$ and $\Phi$ are somewhat similar.
However, as discussed in Section~\ref{sec:results} $c$ is determined from the
turbulent pressure in the 3D simulations, once $\Phi$ has been fixed by 
fitting the frequency location of the depression.}

\begin{figure}
\includegraphics[width=\columnwidth]{\figs/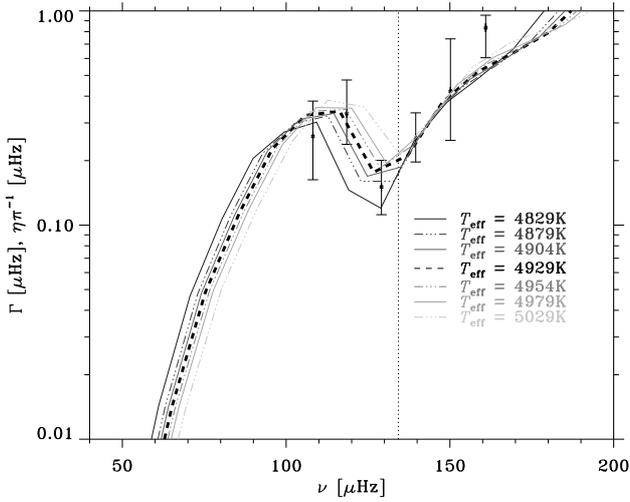}
	\caption{Linewidths from models of KIC 5111718 with different effective temperatures varying $T_{\rm eff}$ in the range $T_{\rm eff,ref}\pm 25,50,100\rm K$.}
	\label{fig:ttest}
\end{figure}

Finally we investigate the effect on modelled linewidths of varying either the mass or effective temperature. The results of the $T_{\rm eff}$ test are shown in Fig.~\ref{fig:ttest}; higher effective temperature systematically increases the linewidths, while lower $T_{\rm eff}$ broadens the trough towards lower frequencies. The relative magnitude of the depression in the damping rates increases with decreasing $T_{\rm eff}$. This is consistent with \Kp\ observations of main-sequence stars \citep{appourchaux2014AA,lund2017ApJ}.

The effects of varying the mass within the uncertainties given by H17, from $1.55\rm M_{\odot}$ to $1.65\rm M_{\odot}$, are only minor.
%

\section{Results}
\label{sec:results}
	Using the methods described in Section \ref{sec:methods} we calculate frequency-dependent damping rates for nine stars in NGC 6819. The stars span an extended part of the RGB evolutionary stage with the most evolved having a large frequency separation of $\Delta\nu = 3.08\rm{\mu Hz}$ and the least evolved $\Delta\nu = 10.50\rm{\mu Hz}$. For most of the stars there is a trough in the damping rates around the frequency of maximum power $\nu_{\rm max}$\footnote{We use the observed values of $\nu_{\rm max}$ given by H17} - a feature which we are able to reproduce in our models. This is also seen for main-sequence solar-like stars; for the Sun, the trough is less deep and runs over five radial orders, \citep[e.g.,][]{houdek2017arXiv}, whereas for the red giants considered here, the interval covers only two or three orders. When calibrating the convection parameters to obtain the best match between theoretical damping rates and observed linewidths, we first determine $c$ as described earlier. The turbulent anisotropy parameter $\Phi$ is then calibrated in order for the depression in the damping rates to occur approximately at the same frequency as for the observations. As Fig.~\ref{fig:ap_dampingrates5111718} shows, changing $\Phi$ can shift the damping rate pattern to lower or higher frequencies. Since $\Phi$ is closely tied to the convective velocities, changing this parameter can also affect the turbulent pressure profile, although the effect on the $p_{\rm t}/p$ peak value is small compared to varying $c$. Whenever $\Phi$ is changed, we also check the correspondence of the turbulent pressure profile with convection simulations and readjust $c^2$ if necessary. Of the remaining two parameters, $a$ predominantly affects only the depth of the depression in the damping rates around $\nu_{\rm max}$ (see Fig.~\ref{fig:aa_dampingrates5111718}), while $b$ to some extent smooths out the depression at $\nu_{\rm max}$ by raising it and decreasing the damping rates at frequencies just below (see Fig.~\ref{fig:ab_dampingrates5111718}). With this in mind, $b$ is first calibrated by matching, as well as possible, the damping rates above and below the depression. Lastly $a$ is calibrated via the magnitude of the depression in the observed linewidths.
{\rd As discussed in Section~\ref{sec:emtest} there is partial degeneracy
between the effects of $a$ and $b$ on the linewidths and consequently some
correlation between the values determined by the fit.
Also, owing to the much lower sensitivity of the linewidths to $b$ the
value of $b$ resulting from the fit is somewhat uncertain.}

It should be noted that, in the following figures, we also plot the errors on the observed frequencies, although they are not visible in the plots, due to their values being of order $\sim 10^{-2}\rm{\mu Hz}$. All stars are single members of NGC 6819 and none are categorized as having experienced non-standard evolution as per H17.

{\rd The quality of the observed linewidths does not support a full statistical
analysis, including a formal $\chi^2$ minimization.
However, in the fitting process we have evaluated $\chi^2$ for several of
the parameter choices; in general, the parameter values listed 
in Table~\ref{tab:models} do correspond to the fit with the lowest $\chi^2$
amongst the parameter sets considered.}

We find a clear tendency for stars with lower $\Delta\nu$, i.e. more evolved, to require an increasingly local solution - specifically $a$ increases drastically with decreasing $\Delta\nu$. Notice also that the five most evolved stars have the highest $\Phi \geq 1.8$. Generally we obtain very good agreement between observed linewidths and modelled damping rates with typically only a single outlying mode for any given star. When calibrating the convection model parameters to obtain a matching model, we focus first and foremost on the modes closest to $\nu_{\rm max}$. In the present section we highlight only a few cases; the results for the remaining cases are shown in Appendix \ref{app:linewidths}.
%
%

\subsection{KIC 5111718}
Fig.~\ref{fig:lw5111718} shows the theoretical damping rates (curves) and observed linewidths (symbols with error bars) as functions of frequency for KIC 5111718. This is the least evolved of the stars considered here with $\Delta\nu = 10.50\rm{\mu Hz}$; also here we find very good agreement between theoretical model damping rates and observations. We highlight this case as it is the model of choice for the tests described in Sections \ref{sec:emtest} and \ref{sec:emtest_results} due to the models being very well behaved with regards to numerical convergence for a wide range of adjustable parameter values.

\subsection{KIC 5111940}
\begin{figure}
\includegraphics[width=\columnwidth]{\figs/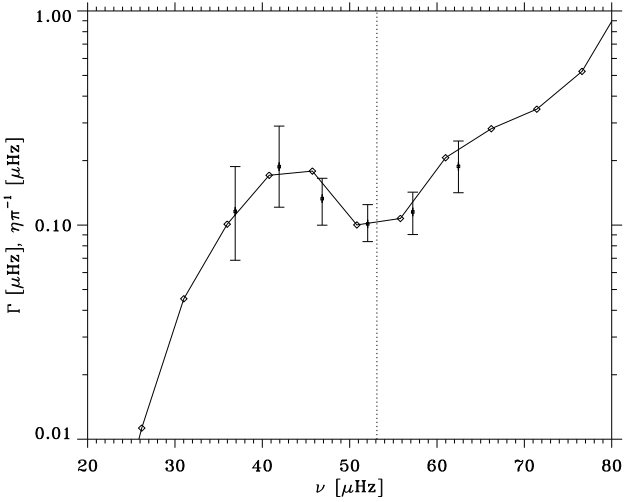}
\caption{Linewidths of KIC 5111940. Diamonds show model computations as $\Gamma = \eta\pi^{-1}$ while the points with error bars are observed linewidths. The dotted vertical line is drawn at the observed $\nu_{\rm max}$.}
	\label{fig:lw5111940}
\end{figure}

Fig.~\ref{fig:lw5111940} shows the frequency-dependent linewidths for KIC 5111940. This is the only case where all the observed linewidths are matched by model damping rates within the observational uncertainties; however the model frequencies are systematically a few $\mu\rm{Hz}$ too low. This happens to be the case for most of the stars seen in Fig.~\ref{fig:HR}. It is tempting to attribute this shift to surface effects, and studies by \cite{trampedach2017MNRAS} do show that the structural surface effect for such an evolved star is comparable to the shift seen in Fig.~\ref{fig:lw5111940} and likewise for the other model results. Our non-adiabatic pulsation analysis accounts for the modal part of the surface effect along with the part of the structural surface effect that is due to turbulent pressure. However, we do not account for the convective expansion of the atmosphere due to convective backwarming, which makes up roughly half of the structural surface effect \citep{RT2010ApSS}. It is an asymmetrical, purely three-dimensional phenomenon that cannot be modelled in 1D \citep{SteinNordlund1998ESASP}. The frequency differences between our models and observations can therefore reasonably assumed to stem from the lack of convective backwarming.

\subsection{KIC 5112734 and 5024583}
\begin{figure}
\includegraphics[width=\columnwidth]{\figs/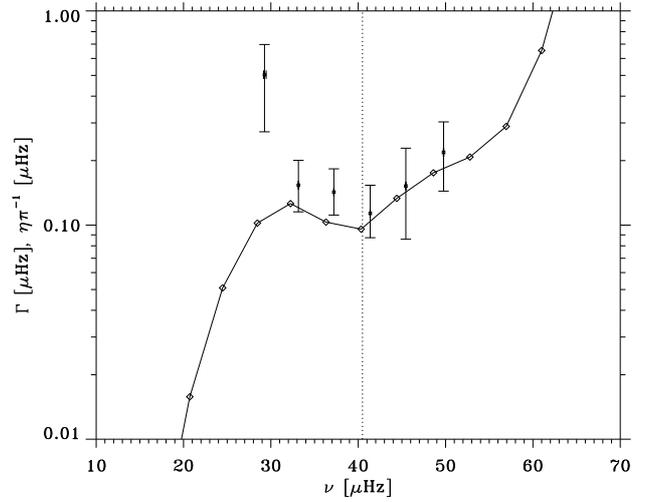}
\caption{Linewidths of KIC 5112734. Diamonds show model computations as $\Gamma = \eta\pi^{-1}$ while the points with error bars are observed linewidths. The dotted vertical line is drawn at the observed $\nu_{\rm max}$.}
	\label{fig:lw5112734}
\end{figure}

\begin{figure}
\includegraphics[width=\columnwidth]{\figs/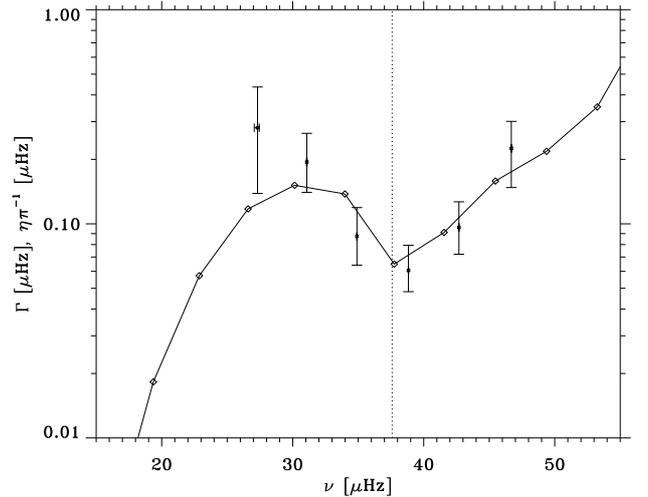}
\caption{Linewidths of KIC 5024583. Diamonds show model computations as $\Gamma = \eta\pi^{-1}$ while the points with error bars are observed linewidths. The dotted vertical line is drawn at the observed $\nu_{\rm max}$.}
	\label{fig:lw5024583}
\end{figure}

It is interesting to compare KIC 5112734 and 5024583. The observed stars are very similar in terms of $\Delta\nu$ and $T_{\rm eff}$ (see Table \ref{tab:models}) but very different in terms of linewidths as evident in Figs~\ref{fig:lw5112734} and \ref{fig:lw5024583}. The former exhibits only a weak depression in the linewidth profile while the latter dips deep at $\nu_{\rm max}$. This is reflected in the convection parameters, where we use $b^2=1200$ for KIC 5024583, which is relatively low for such an evolved star. For the KIC 5112734 model we had to increase both $a^2$ and $b^2$ to very high values in order to reproduce theoretically the depression in the damping rates at $\nu_{\rm max}$.  The depths of the convection zones of the models are $r_{\rm BCZ} = 0.054R$ and $0.053R$, respectively. If the linewidths of these two stars depend on effective temperature and mass in the same manner as does KIC 5111718 (Fig.~\ref{fig:ttest}), then a difference of $20\rm K$ and $0.04\rm M_{\odot}$ does not seem sufficient in explaining the difference in observed linewidth profile between the two.
It speaks to the strength of the method that we are able to reproduce linewidths for a wide range of RGB stars. At the same time, the comparison between KIC 5112734 and 5024583 shows that the physical conditions and dynamics can differ significantly between two stars which appear very similar in their global properties.

\subsection{KIC 5023732 and 5023845}
For the two stars KIC 5023732 and 5023845 we were not able to reproduce successfully the observed linewidths (figures are shown in Appendix \ref{app:linewidths}).
{\rd For neither of the stars the linewidths exhibit the characteristic depression around $\nu_{\rm max}$, which is otherwise a common feature of the observations. 
There is no indication that this pair represents a specific stellar type, for which our model is inadequate.}
Of the stars considered here, KIC 5023732 is the most evolved with $\Delta\nu = 3.08\rm{\mu Hz}$, while KIC 5023845 is one of the least evolved stars analysed by H17 with $\Delta\nu = 8.87\rm{\mu Hz}$. The parameters of the theoretical models are very different as well -- KIC 5023732 having an extremely deep convective envelope, $r_{\rm BCZ} = 0.0445R$, while for KIC 5023845 $r_{\rm BCZ} = 0.0940R$.
{\rd Attempting to fit the observations yielded values of $a$, $b$, and $c$ 
which differ quite substantially between the two stars, although in neither case resulting in a model with damping rates able to match the observed linewidths,
the computed linewidths showing strong depressions.
On this basis we find it extremely likely that the discrepant behaviour is
caused by problems with the observed values.}
%
\begin{table}
\fontsize{0.24cm}{0.3cm}\selectfont
	\centering
	\caption{Key properties and parameter values for stars and their corresponding models. The large frequency separation $\Delta\nu$ is included as an indicator of evolution. The cases are listed from least to most evolved. $a$, $b$, $c$ and $\Phi$ are the convection model parameters appearing in equations \eqref{eq:diffeq_flux}, \eqref{eq:diffeq_grad}, \eqref{eq:diffeq_pt}, and \eqref{eq:Phi}. The nonlocal mixing-length parameter $\alpha_{\rm NL}$ is included for completeness. For uncertainties on observables, see H17.}
	\label{tab:models}
	\begin{tabular}{lcccccccc} 
		\hline
		KIC & $\Delta\nu$ & Mass & $T_{\rm eff}$ & $a^2$ & $b^2$ & $c^2$ & $\Phi$ & $\alpha_{\rm NL}$ \\
		& $[\mu\rm Hz]$ & $[\rm M_{\odot}]$ & $[\rm{K}]$ & & & & & \\
		\hline
		5111718 & 10.50 & 1.60 & 4932 & 900	& 800 & 120 & 1.8 & 2.165\\
 5023845 & 8.87 & 1.61 & 4845 & 800 & 500 & 200 & 1.8 & 2.053\\
 5024405 & 8.25 & 1.43 & 4775 & 800 & 3000 & 110 & 1.7 & 2.057\\
 5024312 & 8.01 & 1.60 & 4816 & 900 & 900 & 200 & 1.8 & 2.077\\
 5024512 & 6.65 & 1.57 & 4826 & 1200 & 5500 & 150 & 2.05 & 2.281\\
 5111940 & 5.14 & 1.62 & 4741 & 1200 & 5500 & 150 & 2.0 & 2.237\\
 5112734 & 4.14 & 1.64 & 4607 & 6000 & 6000 & 300 & 2.0 & 2.075\\
 5024583 & 3.89 & 1.68 & 4627 & 3000 & 1200 & 200 & 1.9 & 2.106\\
 5023732 & 3.08 & 1.60 & 4588 & 3000 & 6000 & 120 & 2.0 & 2.141\\
		\hline
	\end{tabular}
\end{table}
\section{Discussion and conclusions}
We successfully model frequency-dependent damping rates for RGB stars in the open cluster NGC 6819, using a non-adiabatic stability calculation based on a nonlocal, time-dependent convection model. This convection model has already proven successful for solar-like main-sequence stars with a wide range of $T_{\rm eff}$, including examples where the linewidths as a function of frequency are almost constant as shown by \cite{houdek2017arXiv}, modelling damping rates to match observed linewidths from the \textit{Kepler} LEGACY sample \citep{lund2017ApJ}.

In Section \ref{sec:emtest} we perform a thorough test of the model's sensitivity to the convection parameters - some of which can be calibrated via 3D convection simulations. The results fall nicely in line with \cite{Balmforth1992part1MNRAS}, who made a similar test for the solar case. We find that the convection parameter $c$, which controls the non-locality of the momentum flux, is well constrained. As testament to this, we emphasize that the values inferred from convection simulations not only reproduce the maximum value of $p_{\rm t}/p$. The characteristic depression in the damping rates around $\nu_{\rm max}$ coincides as well with that of the observations as shown in Section \ref{sec:emtest_results}. This holds great promise for 3D convection simulations as a calibration source with regards to 1D convection models.

Across the different best fitting models $b$ varies a lot (see Table \ref{tab:models}), in accordance with the fact that its effect on the damping rates is rather weak as per Fig.~\ref{fig:ab_dampingrates5111718}. Note that in Section \ref{sec:emtest_results} $b$ is only varied with reference to one specific value of $a$. Also, when going to very high $b$-values ($b^2>3000$), we approach the local formulation and the effect on $\eta$ diminishes with further increase in $b$. The values given in Table \ref{tab:models} are those that provide the best match between theoretical damping rates and observed linewidths, although one could also obtain fairly good results with an upper limit of, say, $3000$, on $a^2$ and $b^2$. As is the case for $b$, the values given in Table \ref{tab:models} suggest a trend of $\Phi$ increasing with decreasing $\Delta\nu$, yet conclusive statements cannot be made based on the sample size presented here.

One application of our models is the potential for improving the stellar structure in the surface layers of 1D stellar models. Because the mixing-length parameter of the nonlocal envelope model is calibrated to match the full structure evolution model in the deep interiors (see Fig.~\ref{fig:calistruct5111718}), the two models can be patched together for an improved stellar model, where the structure in the outer parts better resembles that of a given real star.

\cite{appourchaux2014AA} found that for main-sequence solar-like stars the amplitude of the linewidth dip around $\nu_{\rm max}$ decreases with increasing effective temperature. The same conclusion was reached by \cite{lund2017ApJ} for the LEGACY stars. The damping rates obtained from varying $T_{\rm eff}$ in our model of KIC 5111718 (Fig.~\ref{fig:ttest}) are consistent with these results. However, if we compare the observed linewidths of all stars modelled here, shown in Section \ref{sec:results} and Appendix \ref{app:linewidths}, there does not seem to be a clear trend between linewidths and effective temperature. Assuming the relation also holds for RGB stars, it is not surprising that we cannot reproduce the dependence on $T_{\rm eff}$, given the narrow range of $T_{\rm eff}$ for our stars and the strong variations in $\log g$ which might be expected to dominate the variation in the properties of the damping rates.

Regarding the relatively high values of the nonlocal convection parameters $a$ and $b$, we gave in Section \ref{sec:emtest_results} a qualitative explanations as to why the value of the convection parameter $a$ appears to increase with evolution. \cite{antoci2014ApJ} applied very similar methods, including using the same nonlocal convection model, to the chemically peculiar $\delta$ Scuti star HD 187547 (using model mass and effective temperature $M=1.85\msun$ and $T_{\rm eff} = 7575 \rm{K}$ respectively). For the nonlocal convection parameters they adopted the values $a^2=b^2=c^2=950$, noting that these values are rather high, tending towards a more local description\footnote{In \cite{antoci2014ApJ} $c^2$ is not calibrated to 3D convection simulations. Its value should be lower and consequently the turbulent pressure would be lower as seen in Fig.\ref{fig:ac_pt}.}. Very similar values were used earlier by \cite{balmforth2001MNRAS} ($a^2=b^2=c^2=1000$), applying the model to roAp stars. The argument for not using a more nonlocal solution (smaller $a$) for the convective heat flux pertains to the shallowness of the surface convection zone of hotter stars. Because the radial extent of the convective envelope is rather small there is less physical reasoning for averaging over several convective eddies. 
It is peculiar that the same convection model, in order to reproduce observed linewidths, requires similar values for stars with either very deep or very shallow convective envelopes but for different reasons. In both cases the arguments are reasonable, which highlights the incompleteness of our current understanding of the interaction between pulsations and turbulent convection. However, it is important to note that the need for predominantly local solutions to the convective heat flux is not a feature of the model. {\rd For cool main-sequence solar-like stars, smaller $a$-values are needed. Houdek et al. (in preparation) have calculated theoretical damping rates to match the observed linewidths of the \textit{Kepler} LEGACY stars. For the cooler main-sequence stars, values of the order of $a^2\sim 100$ are needed to properly reproduce observations, while for the hotter stars the nonlocal convection parameters have large values.
}


The results presented here are the first examples of theoretical, frequency-dependent damping rates in RGB stars able to match high-quality \Kp\ observations. This is an important step towards a better understanding of the dynamics of convection and pulsations in the turbulent envelope regions of these stars.
{\rd Future improvements to the convection model,
including even more extensive use of 3D simulations,
should concern ways to constrain the non-local convection parameters
$a$ and $b$;
in particular, comparing the relatively strong effect of $b$ on the
super-adiabatic gradient (cf.\ Fig.~\ref{fig:ab_grad}) with the results
of simulations may help breaking the degeneracy between $a$ and $b$.
Also, it would be interesting to
incorporate the depth-dependence of the shape factor $\Phi$ determined
from the simulations.
The inclusion} of the kinetic energy flux is currently in the works, its effect on the damping rates being so far unknown.
{\rd The combined effort of improving the modelling in this manner and
extending the analysis to a broader range of observed stars will, one may hope,
improve our understanding of the properties of stellar convection and its
interaction with the pulsations.
Such improvements in the modelling of the near-surface layers
are also crucial for reducing their effect on the computed frequencies and
consequently improving the characterization of stellar properties through
frequency fitting.}

\section*{Acknowledgements}
Funding for the Stellar Astrophysics Centre is provided by The Danish National Research Foundation (Grant DNRF106).



\bibliographystyle{mnras}
\bibliography{references} 




\appendix

\section{Additional figures}
\begin{figure}
\includegraphics[width=\columnwidth]{\figs/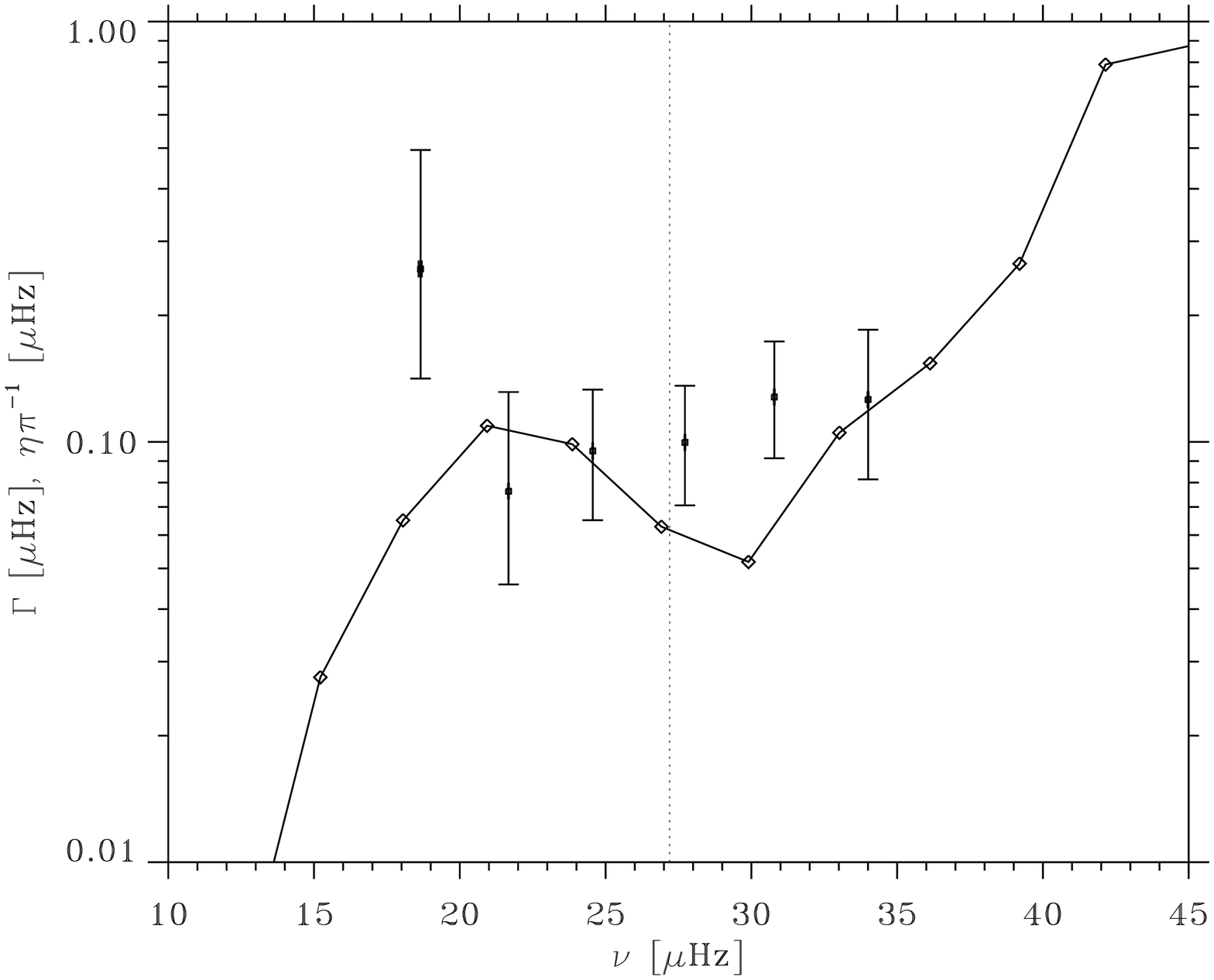}
	\caption{Linewidths of KIC 5023732. Diamonds connected by a full line show model computations as $\eta\pi^{-1}$ while the points with error bars are observed linewidths. The dotted vertical line is drawn at the observed $\nu_{\rm max}$.}
	\label{fig:lw5023732}
\end{figure}

\begin{figure}
\includegraphics[width=\columnwidth]{\figs/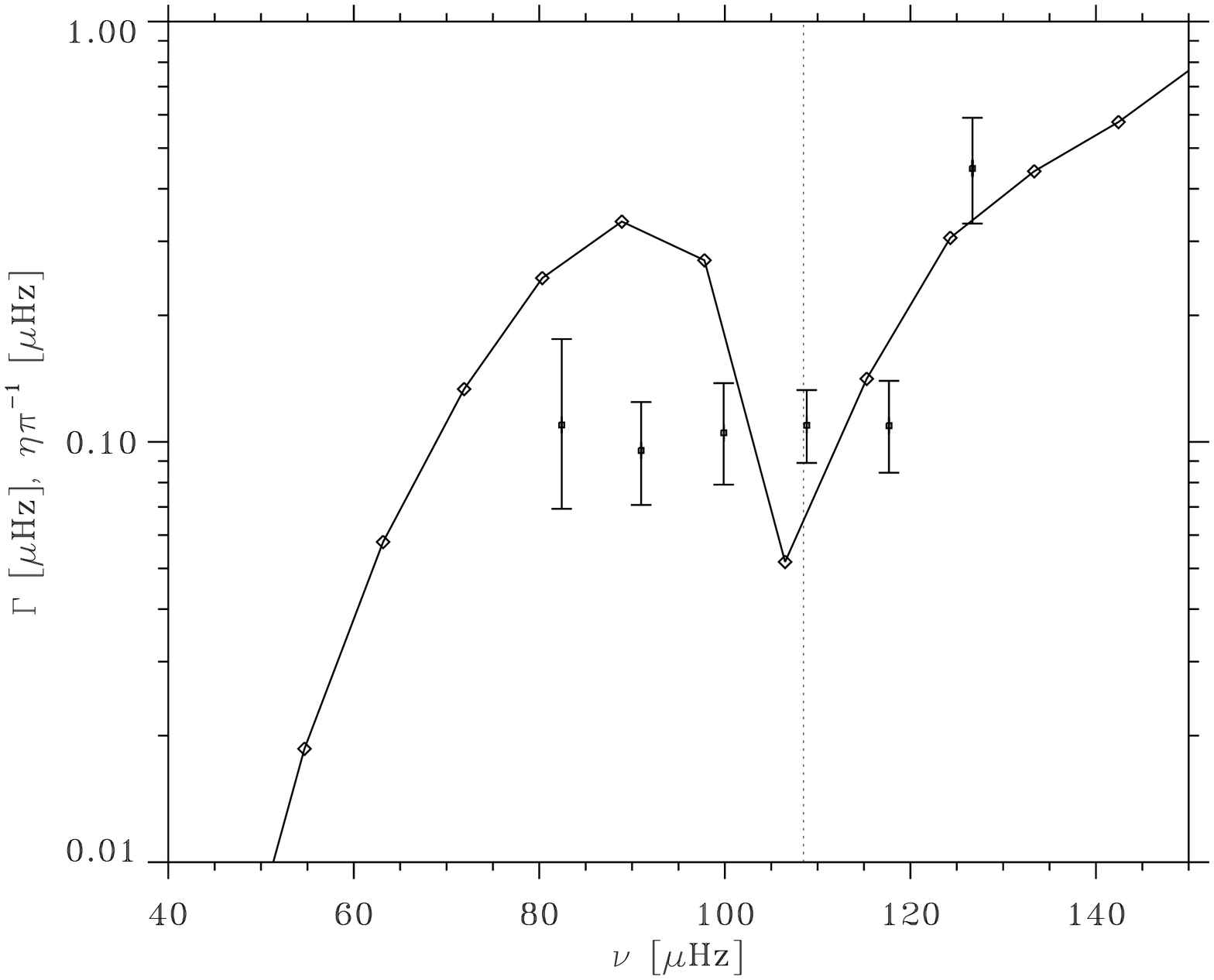}
	\caption{Linewidths of KIC 5023845. Diamonds connected by a full line show model computations as $\eta\pi^{-1}$ while the points with error bars are observed linewidths. The dotted vertical line is drawn at the observed $\nu_{\rm max}$.}
	\label{fig:lw5023845}
\end{figure}

\begin{figure}
\includegraphics[width=\columnwidth]{\figs/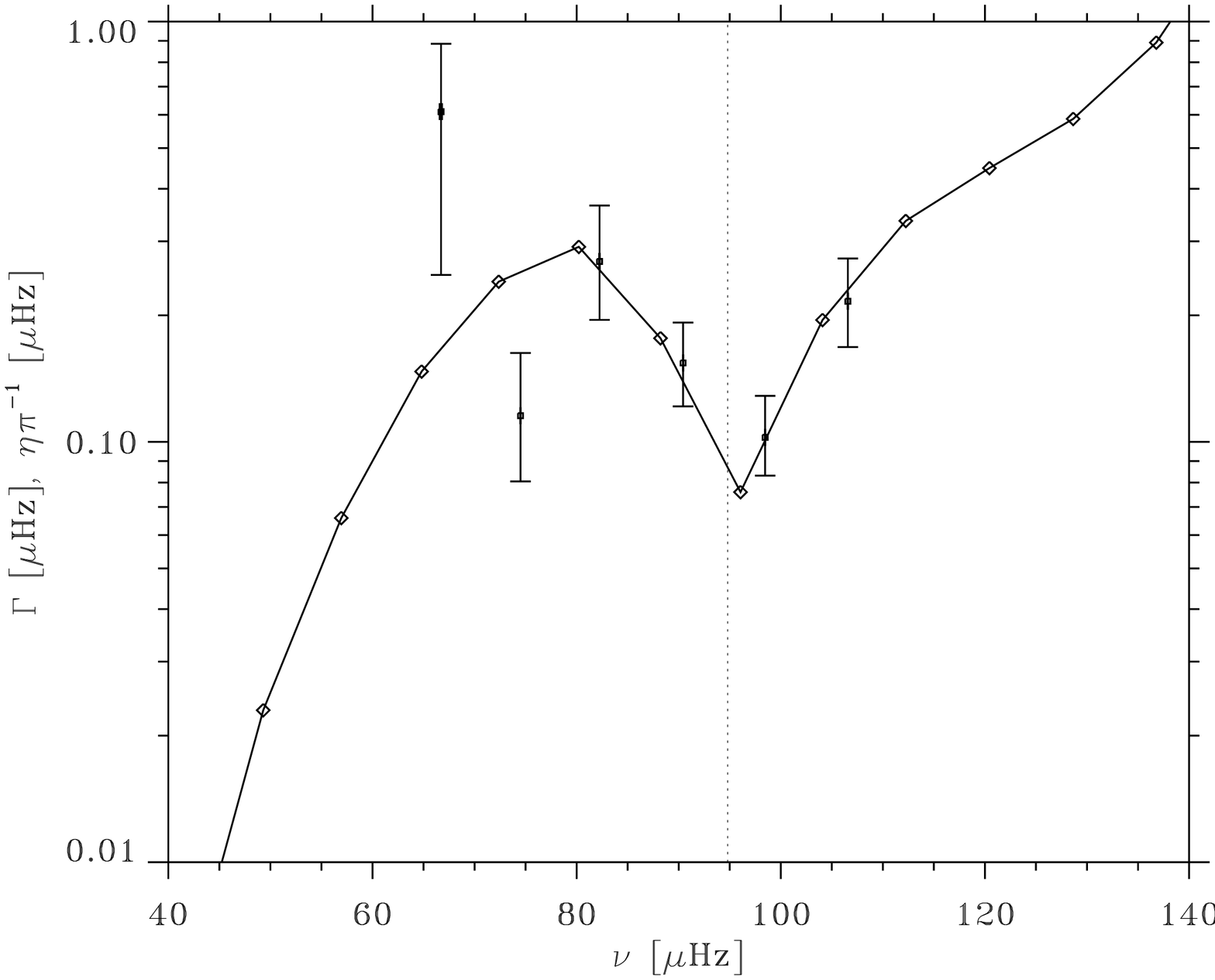}
	\caption{Linewidths of KIC 5024312. Diamonds connected by a full line show model computations as $\eta\pi^{-1}$ while the points with error bars are observed linewidths. The dotted vertical line is drawn at the observed $\nu_{\rm max}$.}
	\label{fig:lw5024312}
\end{figure}

\begin{figure}
\includegraphics[width=\columnwidth]{\figs/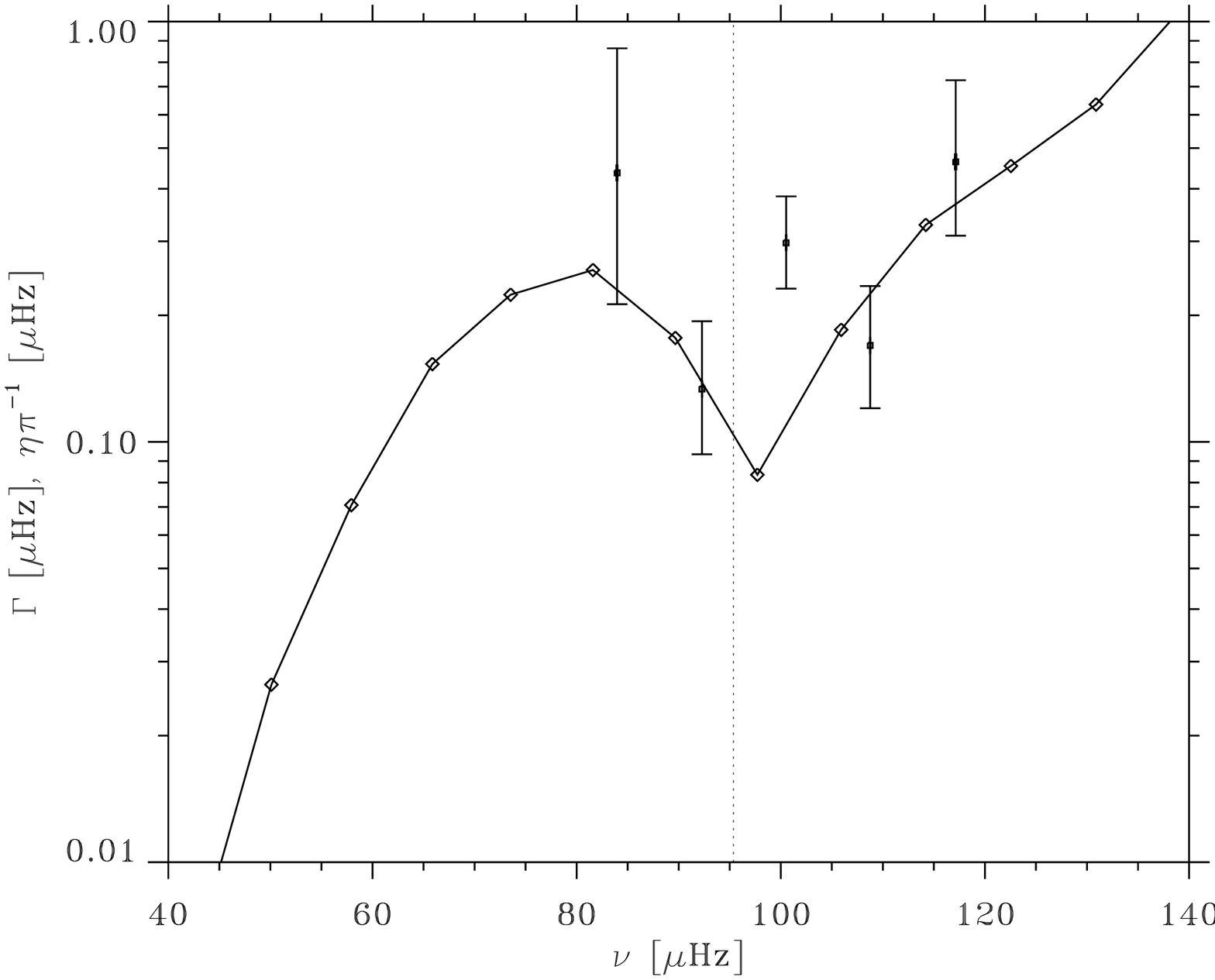}
	\caption{Linewidths of KIC 5024405. Diamonds connected by a full line show model computations as $\eta\pi^{-1}$ while the points with error bars are observed linewidths. The dotted vertical line is drawn at the observed $\nu_{\rm max}$.}
	\label{fig:lw5024405}
\end{figure}

\begin{figure}
\includegraphics[width=\columnwidth]{\figs/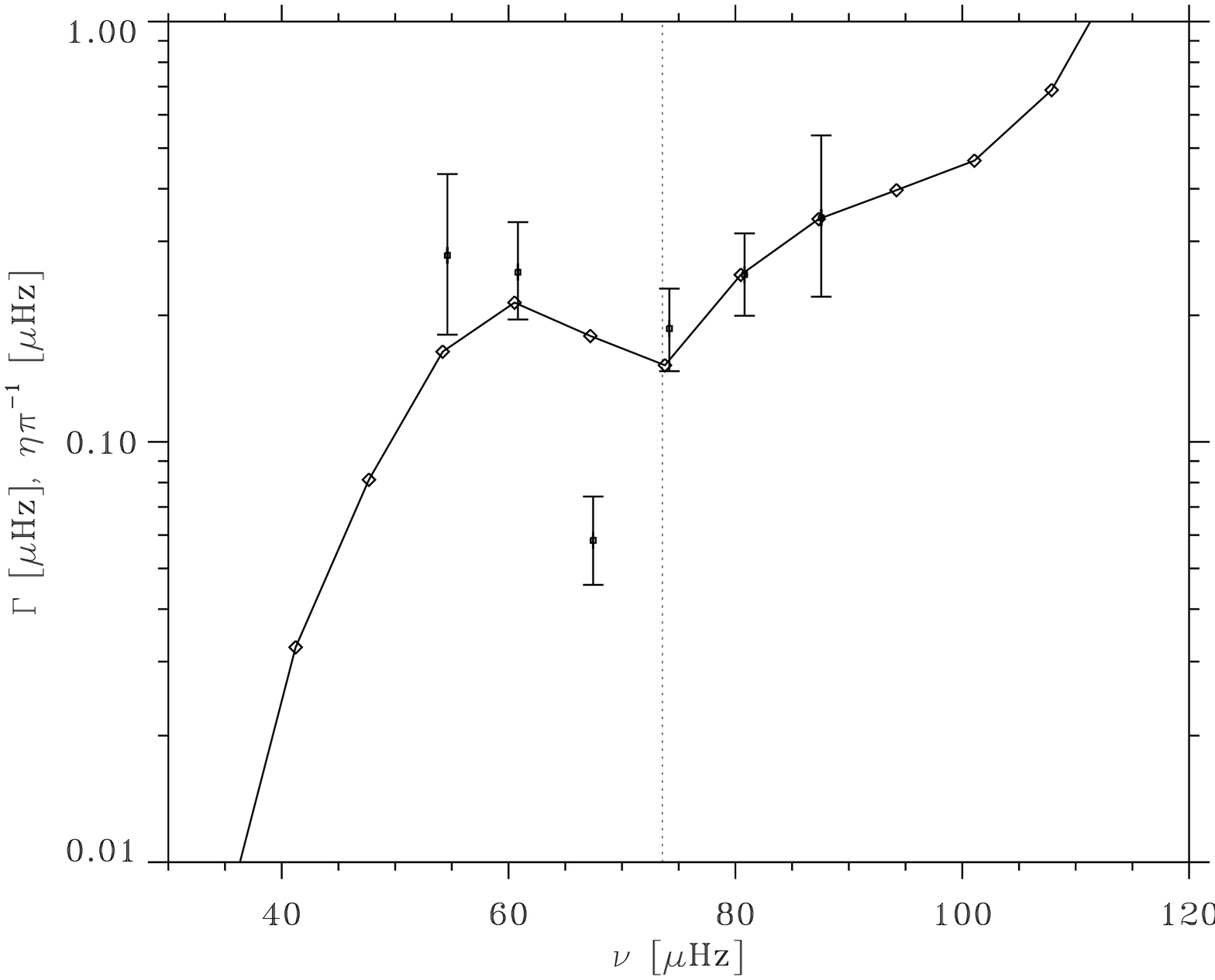}
	\caption{Linewidths of KIC 5024512. Diamonds connected by a full line show model computations as $\eta\pi^{-1}$ while the points with error bars are observed linewidths. The dotted vertical line is drawn at the observed $\nu_{\rm max}$.}
	\label{fig:lw5024512}
\end{figure}
\label{app:linewidths}


\bsp	
\label{lastpage}
\end{document}